\documentclass[11pt]{article}
	
	\newcommand{\blind}{0}
	
    \makeatletter
    \renewcommand\section{\@startsection {section}{1}{\z@}%
                                       {-3.5ex \@plus -1ex \@minus -.2ex}%
                                       {2.3ex \@plus.2ex}%
                                       {\normalfont\fontfamily{phv}\fontsize{16}{19}\bfseries}}
    \renewcommand\subsection{\@startsection{subsection}{2}{\z@}%
                                         {-3.25ex\@plus -1ex \@minus -.2ex}%
                                         {1.5ex \@plus .2ex}%
                                         {\normalfont\fontfamily{phv}\fontsize{14}{17}\bfseries}}
    \renewcommand\subsubsection{\@startsection{subsubsection}{3}{\z@}%
                                        {-3.25ex\@plus -1ex \@minus -.2ex}%
                                         {1.5ex \@plus .2ex}%
                                         {\normalfont\normalsize\fontfamily{phv}\fontsize{14}{17}\selectfont}}
    \makeatother
	\usepackage{amsmath}
	\usepackage{graphicx}
	\usepackage{enumerate}
	\usepackage{xcolor}
	\usepackage{natbib} %comment out if you do not have the package
	\usepackage{url} % not crucial - just used below for the URL
    \usepackage{algorithm}
\usepackage{algorithmic}
\usepackage{enumitem}
\usepackage{xurl}
\usepackage{multirow}
\usepackage{CJKutf8}
\usepackage{caption}
\usepackage{booktabs}
\usepackage{makecell}
\usepackage[utf8]{inputenc}
\usepackage{array} % 用于定义新列类型
\usepackage{tabularx} % 自动调整宽度的表格
\usepackage{setspace} % 控制行间距
\usepackage{amsmath}
\usepackage{adjustbox} % 用于调整表格宽度
\usepackage[T1]{fontenc}
\usepackage{tabularx}
\usepackage{newfloat}
\usepackage{listings}
\usepackage{adjustbox} % 用于调整表格大小
\usepackage{multicol}
\usepackage{csquotes}
\usepackage{subcaption}
\usepackage{subfig}
\usepackage[utf8]{inputenc}
\usepackage[T1]{fontenc}
\usepackage{tcolorbox}
\tcbuselibrary{skins,breakable} % 加载扩展库
\usepackage{adjustbox}
\begin{document}
        \begin{CJK*}{UTF8}{gbsn}
		
			%%%%%%%%%%%%%%%%%%%%%%%%%%%%%%%%%%%%%%%%%%%%%%%%%%%%%%%%%%%%%%%%%%%%%%%%%%%%%%
		\def\spacingset#1{\renewcommand{\baselinestretch}%
			{#1}\small\normalsize} \spacingset{1}
		%%%%%%%%%%%%%%%%%%%%%%%%%%%%%%%%%%%%%%%%%%%%%%%%%%%%%%%%%%%%%%%%%%%%%%%%%%%%%%
		
		\if0\blind
		{
			\title{Rejection or Inclusion in the Emotion–Identity Dynamics of TikTok Refugees on RedNote}
			\author{Mingchen Li $^a$ , Wenbo Xu $^a$, Wenqing Gu $^a$, Yixuan Xie $^a$, Yao Zhou $^a$, Yunsong Dai $^b$\\ Cheng Tan $^{c*}$, Pan Hui $^{a*}$,\\
			$^a$ The Hong Kong University of Science and Technology (Guangzhou), Guangzhou, P.R.C. \\
             $^b$ The School of Business and Management, Jilin University, Changchun, P.R.C. \\
             $^c$ Peking University, Beijing, P.R.C.}
			\date{}
			\maketitle
		} \fi
		
		\if1\blind
		{

            \title{\bf \emph{IISE Transactions} \LaTeX \ Template}
			\author{Author information is purposely removed for double-blind review}
			
\bigskip
			\bigskip
			\bigskip
			\begin{center}
				{\LARGE\bf \emph{IISE Transactions} \LaTeX \ Template}
			\end{center}
			\medskip
		} \fi
		\bigskip
		
	\begin{abstract}
This study examines cross-cultural interactions between Chinese users and self-identified "TikTok Refugees" (foreign users who migrated to RedNote after TikTok’s U.S. ban). Based on a dataset of 1,862 posts and 403,054 comments, we use large language model-based sentiment classification and BERT-based topic modeling to explore how both groups engage with the TikTok refugee phenomenon. We analyze what themes foreign users express, how Chinese users respond, how stances (Pro-China, Neutral, Pro-Foreign) shape emotional expression, and how affective responses differ across topics and identities. Results show strong affective asymmetry: Chinese users respond with varying emotional intensities across topics and stances: pride and praise dominate cultural threads, while political discussions elicit high levels of contempt and anger, especially from Pro-China commenters. Pro-Foreign users exhibit the strongest negative emotions across all topics, whereas neutral users express curiosity and joy but still reinforce mainstream discursive norms. Cross-topic comparisons reveal that appearance-related content produces the most emotionally balanced interactions, while politics generates the highest polarization. Our findings reveal distinct emotion–stance structures in Sino-foreign online interactions and offer empirical insights into identity negotiation in transnational digital publics. \\
	\end{abstract}
			
	\noindent%

	%\newpage
	\spacingset{1.5} % DON'T change the spacing!

\section{Introduction}
Social media platforms have become vital arenas for the production and circulation of cultural symbols. In many countries, these platforms are increasingly integrated into everyday life (\cite{Dhoha2019})—not only as crucial sources of information (\cite{Nielsen2014}) but also as spaces for social interaction, collective expression, and participation in subcultural communities. As such, social media platforms offer an important lens through which to examine a nation’s sociocultural fabric (\cite{Ohiagu2014}). With the deepening of globalization, cross-cultural encounters on digital platforms are no longer limited to transnational networks such as Twitter or Facebook but are also increasingly visible on region-specific platforms like RedNote (Xiaohongshu), a Chinese social media site. Following the ban of TikTok in the United States, a growing number of English-speaking users—self-described as "TikTok refugees" (TtR) (\cite{Yuan2025})—have migrated to RedNote, transforming it into a contested space for cross-cultural engagement. These users often engage in dialogues about bilateral language learning (\cite{Liu2025}), sparking interaction between domestic and foreign users.

While prior research has investigated how Chinese users articulate emotions on RedNote (\cite{Xie2024}) and how foreign users adapt their communication practices to better understand Chinese sociocultural norms (\cite{Zhang2025}) few studies have approached these interactions from a cultural identity perspective. This gap is notable, given that identity-based encounters on social media are often fraught with power asymmetries, symbolic boundaries, and politicized emotions.

Existing literature highlights that transnational social media platforms function not merely as language-learning tools (\cite{Hsiao2014}), but also as arenas of ideological negotiation (\cite{Huang2024}). Cultural identity emerges as both an individual orientation and a collective frame, shaped by algorithmic curation and platform affordances. Research on polarization suggests that echo chambers intensify intra-group cohesion while amplifying inter-group hostility (\cite{Iandoli2021}). In contrast, balance theory posits that platforms with equitable information access and limited exposure to bias may reduce intergroup conflict and promote mutual understanding (\cite{Ludwig2025}). Other work has shown that identity-based tensions are often exacerbated through discursive framings that undermine group legitimacy (\cite{Sparby2017}). Taken together, these findings suggest that both media infrastructure and user-generated content influence cross-cultural affective dynamics.

In the early phase of TikTok refugees’ entry into RedNote, Chinese users often responded with curiosity and openness. However, over time, national histories, geopolitical tensions, and cultural distance began to influence sentiment and interaction patterns (\cite{Choi2011, Guarnieri2025}). Emerging symbolic boundaries—such as language proficiency, physical appearance, and national identity—complicated these exchanges (\cite{Lamont1992}). Attitudes toward foreign users began to vary based on perceived alignment with these identity markers. Thus, digital platforms like RedNote become sites where identity is not only performed but also policed.

This study examines how TikTok refugees’ posts on RedNote elicit affective and discursive responses from Chinese users. We explore not only the thematic content of posts made by English-speaking users but also the emotional and nationalistic responses they generate among Chinese netizens. While existing research has largely examined prejudice and discrimination within the context of U.S. domestic race politics or anger-driven affective publics (\cite{Maxwell2016}), our study introduces an international political lens to examine how contempt, derision, and geopolitical antagonisms are expressed and negotiated online. Prior work in international digital politics has often focused on issues such as platform sovereignty (\cite{Riordan2018}) or the framing of global crises; few studies, however, have analyzed public emotion and identity politics through the lens of everyday platform interactions.

To explore these dynamics, we collected 1,862 notes along with 403,054 corresponding first- and second-level comments. Our final dataset includes 424 notes, authored by 168 individuals (89 foreign and 79 Chinese), and a total of 89,954 comments—45,747 from Chinese authors and 44,207 from foreign authors. These notes cover a broad range of topics and responses, reflecting the diverse perspectives of both foreign and Chinese users. Using large language model-based emotion classification and BERT-based topic modeling, we analyze affective orientation and national attitude in cross-cultural interactions. We offer three key contributions to CSCW:

1.We provide empirical insights into cross-cultural interaction patterns on a Chinese domestic social media platform, contributing a non-Western perspective to the literature on multilingual and transnational platform use.

2.By integrating emotional analysis with political identity theory, we offer a novel framework for understanding how digital publics engage with foreign others through national affect.

3.We theorize the role of identity markers in shaping symbolic boundaries within digital spaces, contributing to ongoing discussions of how platforms mediate cultural belonging and exclusion.

\section{Related Work}
\subsection{Platform Migration and Post-Migration Expression}

Platform migration often reflects both ideological resistance and expressive constraint. Studies such as Zia et al. (\cite{WOS:001353181500010}) on Twitter-to-Mastodon migration and Liu et al. (\cite{liu2025TikTok}) on the influx of TtR into RedNote reveal how users strategically seek platforms that better align with their discursive preferences. Extending this line of analysis, Lutz \& Aragon (\cite{10.1145/3687019}) argue that TikTok’s algorithmic compression of Latinidad identities fuels user movement toward niche communities—paralleling how TtR participants search for less politicized, more personalized digital spaces.

RedNote’s hybrid architecture—blending social commerce and lifestyle discourse—provides a comparatively low-conflict environment for such users. Unlike the confrontational dynamics of Twitter (\cite{WOS:001279591706019}), RedNote fosters casual, informal interactions that lend themselves to affective bonding and soft cultural exchange (\cite{liu2025TikTok}). However, this does not necessarily reduce cultural tension; instead, it channels potential conflict into everyday formats—such as pet images, fashion commentary, or lifestyle posts—that appear apolitical but may still evoke identity-related sensitivities or reinforce symbolic boundaries.

\subsection{Cross-Cultural Emotion and Identity Framing}

Emotional responses in these transnational encounters are not neutral. As prior research indicates, cross-cultural interaction is often shaped by symbolic asymmetries and platform-specific affective norms. The DUT Chinese Sentiment Lexicon (\cite{WOS:001279591706019}) and studies like Foriest et al. (\cite{10.1145/3686940}) on gendered affect demonstrate how emotional vocabularies are stratified by power. For instance, Chinese users' pride in national culture often contrasts with foreign users’ expressions of admiration or cautious respect. This aligns with Wang et al.’s observation (\cite{10.1145/3686898}) that domestic platforms intensify in-group emotional feedback loops—amplifying pride, sarcasm, or contempt in the name of soft nationalism. 

Such emotional asymmetry connects directly to identity performance. RedNote’s hybrid commerce-lifestyle design fosters a form of banal nationalism (\cite{WOS:000071038300004}), distinct from the overt ideological debates on platforms like Twitter. Chinese users’ responses to TtR often embed patriotic narratives, assert cultural authority, and reinforce symbolic boundaries. These replies frequently take the form of subtle cues—such as praise for Chinese traditions, corrective tones, or references to geopolitical achievements (\cite{Shan202125}). As Zhang et al. (\cite{WOS:001279591706019}) argue, Chinese digital spaces encourage "self-mobilized nationalism," which manifests in emotionally charged replies to foreign users. Lou et al. (\cite{10.1145/3710933}) show how anonymity enhances spontaneous collective reinforcement, further amplifying identity superiority through pride and derision.

Conversely, cross-cultural participants may experience implicit forms of exclusion. While racial microaggressions on Western platforms often take overt verbal forms, Gunturi et al. (\cite{10.1145/3637366}) point to cultural gatekeeping and subtle discursive boundaries as RedNote-specific mechanisms of identity policing. These asymmetries—emotional, structural, and symbolic—collectively shape how identity is not only expressed but negotiated in transnational digital publics.

Negative emotions like anger and fear, meanwhile, are not accidental but platform-mediated. This echoes Chancellor et al. (\cite{WOS:000389809500095}) who observe how moderation and platform norms affect the prevalence of lexical affect in pro-ED communities. In RedNote, as we show later, affective expressions of nationalism—whether pride or contempt—are shaped both by user identity and platform architecture.

\subsection{Platformed Affective Norms in Cross-Cultural Interaction}

Existing frameworks of affective polarization (\cite{Ludwig2025}) primarily analyze Western contexts, often overlooking how national cultural scripts and platform-specific emotional norms structure cross-cultural encounters on Chinese platforms. RedNote, for instance, encodes pride-contempt dichotomies not only through algorithmic design—such as the 'positive energy' recommendation logic—but also through culturally resonant discursive expectations that govern user interactions (\cite{2008Facebook}).

Qian et al. (\cite{10.1145/3686980}) illustrate that semi-acquaintance communities on Chinese platforms cultivate socio-emotional support, yet simultaneously reinforce conformity via shared affective expectations. This aligns with pro-China comment patterns, where implicit cultural norms—like downvoting dissent or prioritizing patriotic tone—echo Magalhaes’s notion of "normative regulation" (\cite{WOS:000436866400002}). Similarly, Dym \& Fiesler’s bottom-up approach(\cite{WOS:000555422000014}) highlights how top-level comments on TtR posts often set emotional and ideological baselines for replies, reinforcing platform-specific expressions of national belonging.

These culturally embedded emotional scripts play a key role in shaping how Chinese users navigate and respond to foreign presence, particularly in the context of identity assertion and affective alignment.

\section{Critical Gaps and Theoretical Innovation}

While prior work examines migration (\cite{WOS:001353181500010}) and nationalism (\cite{Shan202125}), few studies analyze cross-cultural affective feedback loops on non-Western platforms. Our focus on temporal affective shifts bridges Gunturi et al.’s (\cite{10.1145/3637366}) linguistic analysis and Foriest et al. (\cite{10.1145/3686940})’s muting studies. By contrasting RedNote with Twitter, we theorize how platform-specific affective norms and interactional structures shape expressions of national identity and reinforce symbolic boundaries in transnational encounters.

In summary, while prior work has examined cross-cultural platform migrations (e.g., Twitter to Mastodon), little is known about how geopolitically displaced communities like TikTok Refugees negotiate identity and belonging on regionally dominant platforms such as RedNote (Xiaohongshu). To address this gap, this study proposes 4 questions:
\textbf{RQ1: What discursive strategies and topics do foreign users (TikTok Refugees) employ when engaging with the TikTok ban and Sino-American relations on RedNote, and how do these reflect their self-presentation as cultural migrants?}

Next, although studies have analyzed nationalist sentiment in Chinese digital spaces, the reciprocal dynamics between foreign and domestic users—particularly how Chinese netizens frame their responses to migrant communities—remain underexplored. Building on theories of affective polarization, this study investigates:
\textbf{RQ2: How do Chinese users respond to TikTok Refugees’ posts, and to what extent do these replies reinforce or challenge narratives of Chinese superiority (e.g., "China Wins" rhetoric)?}

Furthermore, while platform-mediated identity performance has been studied in isolation, the relational alignment of stances (Pro-China, Neutral, Pro-Foreign) between foreign and domestic users remains unclear. Thus, this study asks:
\textbf{RQ3: How do the expressed stances of TikTok Refugees and Chinese users diverge or converge across topics (e.g., politics, culture), and what does this reveal about RedNote as a contested space for cross-cultural position-taking?}

Finally, although emotional expression in transnational online spaces has been attributed to algorithmic bias, the asymmetries in how foreign and domestic users perform emotions (e.g., pride vs. contempt) under shared geopolitical tensions are poorly understood. This study therefore wonders:
\textbf{RQ4: How do the emotional profiles (e.g., pride, anger, contempt) of TikTok Refugees and Chinese users differ, and how do these differences reflect RedNote’s role in amplifying or suppressing specific affective norms in cross-cultural encounters?}

\section{Dataset}

The user interface of Rednote closely resembles Instagram, where each \textit{note} functions similarly to a post. Interaction occurs across various layers of the comment section, including root comments, first-level, and second-level replies. Users frequently include hashtags in the title and content of notes, which serve as important indicators for identifying relevant topics.

We collected data from February to April 2025 using an open-source scraping project\footnote{\url{https://github.com/NanmiCoder/MediaCrawler}}. Based on a predefined keyword list (see Table~\ref{tab:crawling_keywords}), we initially gathered 11,181 related notes. We then extracted their corresponding comments using the Bazhuayu platform\footnote{\url{https://www.bazhuayu.com/}}. Due to many notes having few comments, the final collection included 1,862 notes and 403,054 comments. After preprocessing the text (removing URLs, mentions, hashtags, and emojis), 332,642 comments remained. We manually excluded notes unrelated to the ``TikTok refugee'' topic or with fewer than three comments, resulting in a final dataset of 429 notes and 209,685 comments.

Bazhuayu simulates user behavior by scrolling from top to bottom when accessing note pages. As a result, the comments were stored in the same sequential order as they appear on the interface—root comment, followed by first-level and then second-level replies. This structure preserves the contextual flow, which is beneficial for downstream text analysis.

The user profile page on Rednote also closely resembles that of Instagram. To explore user responses to the TikTok refugee event, we also collected metadata on user profiles. The final dataset includes 377 note authors. For these users, we collected additional information including IP location, profile tags, bio, number of followers, likes received, and titles of all notes they had posted.

After filtering out five categories of notes and their comments, our dataset includes 424 notes, 168 note authors, with 89 being foreign and 79 being Chinese. There are 89,954 comments in total, with 45,747 comments in the sections of Chinese authors and 44,207 comments in those of foreign authors.

To protect user privacy, we anonymized the dataset by directly removing user-identifiable fields such as user IDs, usernames, signatures, and avatars. We also removed IP location information and truncated timestamps to the day level.

\begin{table}[htbp]

\centering
\caption{List of Keywords Related to TtR}
\label{tab:crawling_keywords}

\begin{tabular}{|c|l|}
\hline
\textbf{Index} & \textbf{Keyword} \\ \hline
1 & TikTok\_refugee \\ \hline
2 & TikTokrefugee \\ \hline
3 & ttrefugee \\ \hline
4 & hello china \\ \hline
5 & TikTok \\ \hline
7 & TikTok REFUGEE \\ \hline
8 & cattax \\ \hline
9 & america \\ \hline
10 & TikTokban \\ \hline
11 & fyp \\ \hline
\end{tabular}
\end{table}

% Rednote的UI与Instagram十分类似，note类似instagram的post，评论区的母评论、一级评论和二级评论之间也会有互动。rednote用户会在note的title和content里加上不同的hashtag，也可以作为判断和研究主题相关性的依据。

% 数据集的爬取周期是2025年2月-2025年4月，我们先用open-source scraping project\footnote{https://github.com/NanmiCoder/MediaCrawler},根据keyword list showed in Appendix Table \ref{tab:crawling_keywords}爬取了11181个相关的notes，再用\footnote{https://www.bazhuayu.com/}爬取笔记的评论。因为许多笔记评论数较少，我们总共爬取到了1862条笔记和对应的403054条评论。接着，我们进行了数据预处理，removing URLs, mentions (e.g. @username), hashtags, and emojis，剩下了332642条评论。我们人工去除了与TikTok难民无关或评论数过少（小于3）的笔记，最后剩下429个笔记和209685条评论。因为八爪鱼模拟用户浏览爬取笔记网页时从上往下滑动，因此各级评论在数据存储上也遵循网页浏览顺序，即一行母评论后面紧跟着一级评论，一级评论后紧跟着二级评论，如此循环往复。这也为我们后续的文本分析提供了丰富上下文。

% 为了探究不同用户对TikTok refugee事件的反应，我们还爬取了部分用户信息。我们最终数据集包含了377个笔记作者，以及146973个评论发布者。为了分析更有代表性的用户行为模式，我们从评论发布者中筛选出了407发布评论数大于6条的评论者。利用这些信息，我们爬取了用户的IP属地、标签、简介、粉丝数、获赞量和发布的所有笔记的标题。

\section{Methodologies}
\subsection{Notes Labeling Topic Modeling}

% 我们对424个笔记的主题进行了人工分类，分为：对TikTok refugee事件本身的讨论、政治、社会、文化、颜值、生活、宠物。5位来自社会科学、政治科学和计算机科学的研究者共同标注，卡帕值为0.89。每个分类对应的笔记数量及比例，以及这些笔记对应的评论数量及比例are shown in Table \ref{topic_counts}.

We manually categorized the topics of 424 notes into the following categories: discussion of the TikTok refugee event itself, politics, society, culture, appearance (looks), lifestyle, and pets. The coding was performed by five researchers from the fields of social sciences, political science, and computer science, with a Kappa value of 0.89. The distribution of notes and comments across categories is shown in Table \ref{tab:topic_counts}.

\begin{table}[h]
\centering
\caption{Topic Counts and Proportions for Posts and Comments}
\adjustbox{width=\textwidth}{
\begin{tabular}{ccccc}
\toprule
Topic & Comments Count & Comments Proportion & Notes Count & Notes Proportion \\
\midrule
Culture & 30333 & 0.329911 & 84 & 0.198113 \\
TikTok Refugee Phenomenon & 27831 & 0.302698 & 51 & 0.120283 \\
Society & 16463 & 0.179057 & 36 & 0.084906 \\
Appearance Evaluation & 11834 & 0.12871 & 19 & 0.044811 \\
Politics & 5482 & 0.059624 & 13 & 0.03066 \\
Lifestyle & 82806 & 0.394907 & 163 & 0.415094 \\
Pet & 19856 & 0.094694 & 58 & 0.136792 \\
\bottomrule
\label{tab:topic_counts}
\end{tabular}}
\end{table}

After manually annotating the notes, we conducted topic modeling on the comments under each category. We employed BERTopic (\cite{grootendorst2022bertopic}) along with TopicTuner\footnote{https://github.com/drob-xx/TopicTuner}, which helped optimize the parameters for "min cluster size" and "sample size". The model was trained on the entire comment dataset and separately applied to category-specific subsets. To capture nuanced differences across categories, we used consistent parameters: a minimum cluster size of 30 and a minimum sample size of 3.

\subsection{Sentiment Analysis}

\subsubsection{Technical Implementation}

In scenarios with limited or no labeled samples, large language models have demonstrated superior performance in sentiment analysis tasks compared to smaller language models.(\cite{zhang-etal-2024-sentiment}) Additionally, large language models are capable of leveraging chain-of-thought reasoning to better understand implicit sentiment by incorporating contextual information.(\cite{fei-etal-2023-reasoning}) Therefore, this paper employs the large language model GPT-4o as the core tool for conducting sentiment analysis on comment texts and incorporates chain-of-thought to enhance the analysis. To comprehensively capture the potential emotional dimensions expressed in the comments, we have designed a multi-emotion classification framework encompassing ten common emotion categories: Contempt, Jealousy, Disgust, Fear, Anger, Surprise, Praise, Pride, Joy, and Respect. For each emotion category, we utilize a five-point Likert scale for scoring, with options including "Strongly Disagree," "Disagree," "Neither Agree nor Disagree," "Agree," and "Strongly Agree." This scoring mechanism effectively quantifies the intensity and orientation of emotions expressed in the comments.

In practical applications of GPT-4o for sentiment analysis, we employed the DSPy framework to implement chain-of-thought reasoning and ensure the standardization and consistency of the outputs. DSPy facilitates the design of declarative structured prompts, ensuring that the model’s outputs strictly adhere to predefined formatting requirements.(\cite{dspy})

Furthermore, regarding the parameter settings of the large language model, existing studies suggest that lower temperature settings tend to enhance the determinacy and reproducibility of the model's outputs.(\cite{herrerapoyatos2025overviewmodeluncertaintyvariability}) Consequently, we configured the temperature parameter to 0.1.

\begin{figure}[htbp] % [htbp] 是浮动参数，控制图片位置
    \centering % 图片居中
    \includegraphics[width=0.6\textwidth]{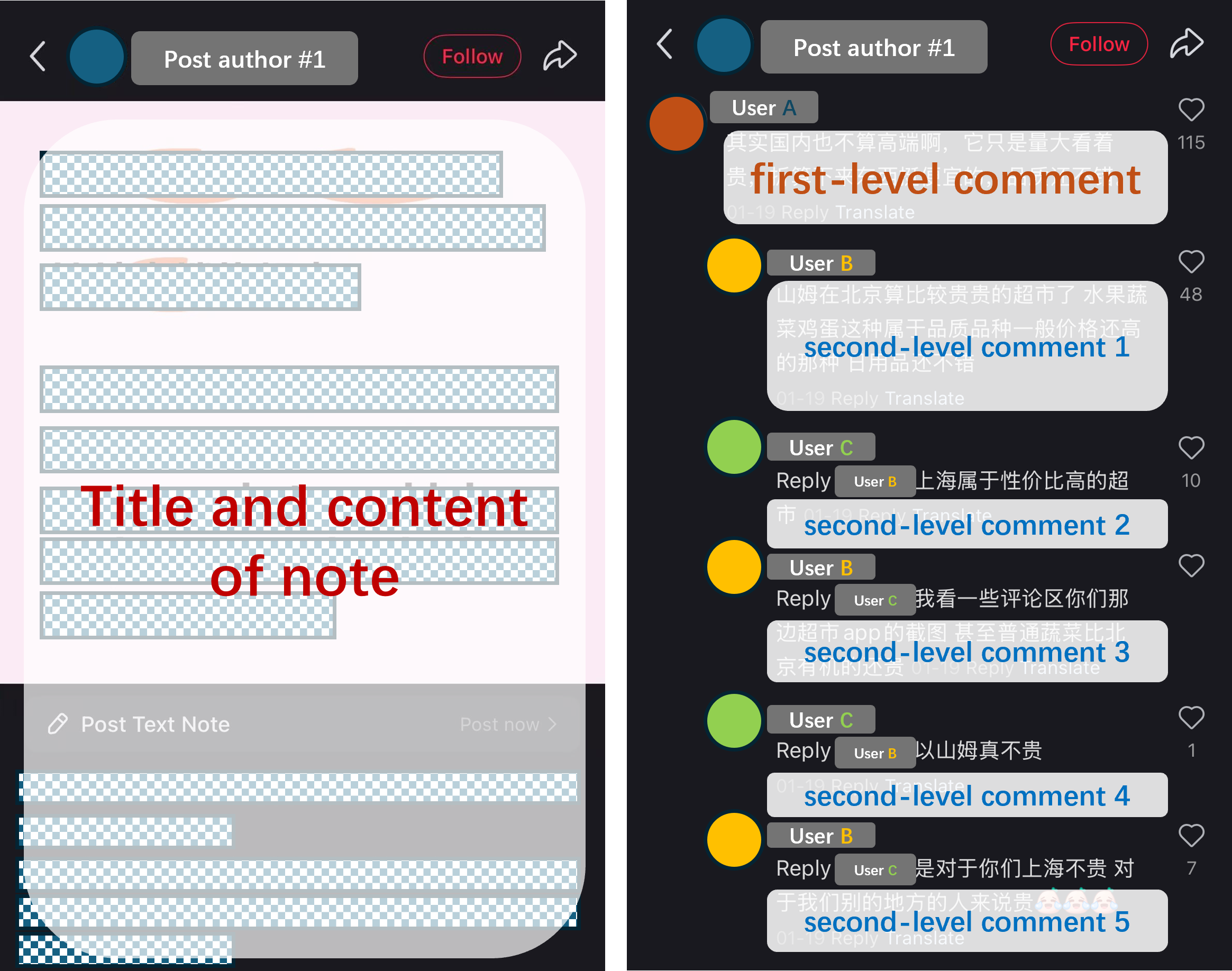} % 插入图片并设置宽度
    \caption{The Comment Section and Contextual Structure of RedNote} % 图片标题
    \label{reply_context} % 标签，用于引用
\end{figure}

% 大语言模型情感分析的技术实现
%% 由于在少量或无标记样本的情况下，大语言模型在情感分析任务上的表现均优于smaller language models。并且大语言模型能够利用思维链，结合语境信息更好的进行implicit sentiment理解。因此，本文在对评论文本进行情感分析时，采用了大语言模型GPT-4o作为核心工具。为了全面捕捉评论中可能表达的情感维度，我们设计了一套多情感分类框架，涵盖10种常见情感类别：Contempt, Jealousy, Disgust, Fear, Anger, Surprise, Praise, Pride, Joy, Respect。对于每种情感类别，我们使用五级Likert量表进行评分，其选项分别为“强烈反对”、“不同意”、“既不同意也不反对”、“同意”和“强烈同意”。这一评分机制能够有效量化评论中情感的强度与倾向。

% 在实际使用GPT-4o进行情感分析时，为了实现思维链，并确保输出结果的标准化和一致性，我们采用了DSPy框架实现思维链和结构化输出。DSPy通过声明式结构化提示词的设计，使得模型生成的输出能够严格遵循预定义的格式要求。

% 此外，在大语言模型的参数设置上，由于在较低的温度设置下，大语言模型的输出会更加有确定性和可重复性。因此，我们采用了temperature=0.1。

\subsubsection{Contextual Framework}

Due to the comment section mechanism of the RedNote research platform, analyzing the semantic information of many comments without their context may result in incomplete interpretations. Therefore, to ensure that sentiment analysis is conducted within the appropriate context, we have designed a Contextual Framework to assist large language model in better understanding the emotional content of comments.

Specifically, the Contextual Framework first determines the hierarchical structure of the comments. For top-level comments (direct replies to posts), the context is composed of the post and its title being replied to. For second-level comments (replies to top-level comments), the context includes the corresponding top-level comment, the post, and its title. Additionally, RedNote’s second-level commenting mechanism allows users to reply to other second-level comments under the same top-level comment, further increasing the complexity of the context. To address this, our Contextual Framework also incorporates the nickname of the user being replied to and their second-level comment. If the second-level comment itself replies to another comment, the framework recursively includes those replied-to comments in the context. To prevent excessive recursion from leading to high computational complexity, we limit the maximum recursion depth to 10 levels. This design ensures sufficient contextual information while maintaining computational efficiency.

% 前后文机制
% 由于研究平台RedNote的评论区机制，许多评论在脱离上下文的情况下分析其语义信息可能是不完整的。因此，为了确保情感分析能够在正确的语境下进行，我们设计了一种前后文机制（Contextual Framework），以帮助大语言模型更好地理解评论的情感内容。

% 具体而言，前后文机制首先判断评论的层级结构。对于一级评论（直接回复帖子的评论），其上下文由所回复的帖子及其标题构成；对于二级评论（回复一级评论的评论），其上下文则包括对应的一级评论、帖子及其标题。此外，RedNote的二级评论机制允许用户针对同一条一级评论下的其他二级评论进行回复，这进一步增加了上下文的复杂性。为了应对这种情况，我们的前后文机制还包含了被回复的用户的昵称以及他们的二级评论，并且如果该二级评论本身还回复了其他评论，则需要递归式地将这些被回复的评论纳入上下文中。为避免大量的递归导致计算复杂度过高，我们限制递归深度最多为10次。这种设计在保证足够语境信息的同时，也兼顾了计算效率。

As illustrated in Figure \ref{reply_context}, taking "second-level comment 5" as an example, the contextual information for this comment includes the "Title and content of note," the "first-level comment," and "second-level comments 1-4." This is because second-level comments 1-4 form the reply chain to which second-level comment 5 belongs.

% 如图\cite{reply_context}所示，以“second-level comment 5”为例，该评论的语境信息包括了图中所示的“Title and content of note”，“first-level comment”，以及“second-level comment 1-4”，这是由于second-level comment 1-4构成了second-level comment 5的回复链。

The prompt used for sentiment analysis with the large language model is as follows.

% sentiment analysis使用的大语言模型prompt如下。

\begin{tcolorbox}[title=Prompt]
    \textbf{}You are required to analyze the comments under posts on RedNote and analysis the sentiment orientation of the comments with respect to the discussed topic.

    \textbf{}Please take into account the context, including the content of the post, the comments being replied to, and the relationships between comments. Based on the following sentiment categories (Contempt, Jealousy, Disgust, Fear, Anger, Surprise, Praise, Pride, Joy, Respect), evaluate the sentiment expressed in each user comment.

    \textbf{}The scoring is divided into five levels, as follows. Please respond with a number from 1 to 5 to indicate the level of sentiment intensity:
    
    \textbf{}1: Strongly disagree
    
    \textbf{}2: Disagree
    
    \textbf{}3: Neither agree nor disagree
    
    \textbf{}4: Agree
    
    \textbf{}5: Strongly agree

    \textbf{}The following is the comment you need to analyze and its context information.

    \textbf{}"Post title: \{Post\_title\}"

    \textbf{}"Post content: \{Post\_content\}"

    \textbf{}"User nickname: \{username\}"

    \textbf{}"The content replied to in this comment: \{replies\_context\}"

    \textbf{}"comment: \{comment\}"
\end{tcolorbox}

\subsection{Stance Detection}

For stance detection, we adopt the same technical approach as used in sentiment analysis. In terms of predefined stance categories, we classify stances into three types: "Pro-China," "Pro-Foreign," and "Neutral."

% 对于stance detection，我们采取了与sentiment analysis相同的技术方法。在立场类别的预定义上，我们将立场分为三类，分别为“偏向中国”、“偏向外国”、“中立”。

The prompt used for stance detection with the large language model is as follows.

\begin{tcolorbox}[title=Prompt]
    \textbf{}You are required to analyze the comments under posts on RedNote and detect the stance expressed in each comments.

    \textbf{}Please take into account the actual situation, considering the contextual information (such as the content of the post, the comments being replied to, and the relationships between comments). Based on a comprehensive analysis of the background information and the comment itself, classify the stance of the comment into one of the following categories: Pro-China, Pro-Foreign, or Neutral.

    \textbf{}The following is the comment you need to analyze and its context information.

    \textbf{}"Post title: \{Post\_title\}"

    \textbf{}"Post content: \{Post\_content\}"

    \textbf{}"User nickname: \{username\}"

    \textbf{}"The content replied to in this comment: \{replies\_context\}"

    \textbf{}"comment: \{comment\}"
\end{tcolorbox}

To ensure accuracy, three researchers manually reviewed a sample of 2,000 comments, proportionally selected based on stance. The Kappa value was 0.88. The detection achieved an F1-score of 79.52\%, with an accuracy of 79.34\%.

\subsection{User Identity Detection}

Since a user's IP location often cannot accurately reflect their identity, and given that some individuals with a Chinese cultural background in the RedNote community may "disguise" themselves as members of the TtR culture to gain higher post visibility, we employ GPT-4o for user identity detection. Specifically, we comprehensively analyze multiple factors, including the user’s nickname, IP location, profile tags, bio, and the titles of all notes they have published, to classify their identity. Based on their cultural background, users are categorized into two groups: those with a Chinese cultural background and those with other cultural backgrounds. The group with a Chinese cultural background includes overseas Chinese and non-Chinese nationals of Chinese descent, while the group with other cultural backgrounds also encompasses foreigners residing in China.

% 由于用户的IP属地在很大程度上不能准确的反映用户的identity，同时在RedNote社区中，存在一些中国文化背景的人为了发布的帖子获得更高的浏览次数而“伪装”成TtR的情况，因此，对于user identity detection，我们使用大语言模型，综合分析用户的昵称，IP location, profile tags, bio和发布的所有笔记的标题对用户的identity进行分类。基于用户的文化背景，我们将用户分为两类，一类是中国文化背景的人；另一类是其他文化背景的人。其中，中国文化背景的人还包括在海外的中国人和非中国国籍的华人，同理，其他文化背景的人也包括旅居中国的外国人。

The prompt used for user identity detection with the large language model is as follows.

\begin{tcolorbox}[title=Prompt]
    \textbf{}You are required to perform identity recognition on user comments sourced from RedNote, categorizing these users into two groups: "Chinese" and "Foreign."
    
    \textbf{}The definition of a "Chinese" user is: A user with a Chinese cultural background (including overseas Chinese).
    
    \textbf{}The definition of a "Foreign" user is: A user without a Chinese cultural background (including foreigners residing in China).
    
    \textbf{}Please take the following information into comprehensive consideration and infer the identity of each user.

    \textbf{}"User nickname: \{username\}"

    \textbf{}"User IP location: \{User\_IP\}"

    \textbf{}"profile tag: \{profile\_tag\}"

    \textbf{}"User Profile: \{bio\}"

    \textbf{}"Titles of all notes they had posted: \{titles\}" (It may contain the titles of multiple posts)
\end{tcolorbox}

% 在筛选出5类笔记及其评论后，我们的数据集中有168个笔记作者，外国作者89个，中国作者79个。中国作者的评论区有45747条评论，外国作者的评论区有44207条评论。

\section{Results}

\subsection{Topic Analysis}
After topic modelling, we divided the discussion into six topics. The initial topic labels were: pets, politics, society, life, culture, and appearance evaluation.

Notably, all six topics often emerge in users’ responses to “greeting”-type posts. After exchanging pleasantries, Chinese and foreign users frequently transition into discussions of everyday life details, which then expand into comparisons of political systems, social structures, and cultural differences. This pattern of interaction lays the groundwork for the development of subsequent thematic discussions. The following analysis excludes the themes of "Life" and "Pets" due to their limited discriminative power in cross-cultural topic modelling.

\subsubsection{Pet topic}

This topic, represented by the phrase "cat tax," focuses on how users establish connections with existing platform members by posting pet photos, especially of cats. Frequently appearing keywords include “kitty,” “cat,” “kitten,” “spotted cat,” “this cat,” “cute,” “my pet,” “puppy,” “pet,” “doggy,” “cat lover,” “owner,” and “dog,” suggesting that both posts and comments are not limited to cats but also involve a broader trend of sharing pet images.

In the context of RedNote, “cat tax” refers to the strategic act of attaching a cute photo of one’s pet—usually a cat—when posting content, in order to gain attention, receive likes, or ease communication. After TikTok refugees and other outside users joined RedNote, they frequently used tags like “paying cat tax” or “my cat is watching me” to quickly integrate into the community by leveraging the universal appeal of pets. Representative examples such as “cat tax: cat outfit” or “cat tax: cute cat” combine pet imagery with content about fashion and lifestyle.

In the comment section, users often respond with emotional expressions such as “cute,” “soft,” “healing,” or “so well-behaved,” creating a warm, humorous, and relaxed atmosphere. Due to their strong visual appeal and emotional resonance, pet images serve as a bridge for non-confrontational interactions between TikTok refugees and local users. This also reflects a broader social strategy where users in cross-cultural settings use pet-related content to establish common interests and reduce communication barriers.

\subsubsection{Political Topic}
This topic focuses on how users express their views and engage in discussions related to national policies, international relations, and social governance in the comment sections. Representative keywords include “China,” “United States,” “policy,” “democracy,” “visa,” “war,” “sanctions,” and “freedom.” These high-frequency terms often co-occur with specific events such as China–U.S. relations, the China–India border conflict, and visa restrictions.

User comments under such posts reflect a range of positions. Some support the government’s stance, with remarks like “A strong nation gives us confidence” or “Patriotism is not a crime,” while others express skepticism or criticism, such as “Where is freedom of speech?” or “The cost of immigration is too high.” Comments often include references to official news sources, map annotations (e.g., “Diaoyu Islands belong to China”), or hashtags (e.g., “\#SanctionUSImports\#”) to reinforce political positions.

\subsubsection{Society topic}
The society topic centers on discussions between TikTok refugee users and Chinese-speaking RedNote users regarding economic development and individual living conditions under the different social systems of China and the United States. This topic often originates from casual “greeting” posts, where users shift from initial pleasantries to deeper conversations about real-world social issues. Common topics include employment, income, rent, parental leave, childbirth, education, healthcare, overwork, social pressure, difficulties in buying property, and family planning policies. These conversations reflect a collective sense of concern about the cost of living and social inequality in the United States.

Users from both sides often combine personal experience with broader structural critique. Some comments share real-life struggles and practical advice, such as “we are all just workers” or “you need to save a 0.5 down payment to buy a home.” Others offer comparative critiques of the two countries’ social systems, pointing to problems like inadequate social security or unequal access to education. Vocabulary related to wages, taxation, visa systems, social class, and family roles further highlights how users understand and construct their positions within different societal structures.

In addition, some users bring in topics such as holidays, consumer goods, and technology, revealing how lifestyle issues intersect with structural concerns. Overall, this topic shows how users turn to online communities for empathy and support while engaging in comparative reflection on governance models in both countries—mostly through descriptive rather than overtly ideological language.

\subsubsection{Life Topic}
The "Life" theme centers on cross-cultural everyday conversations between TikTok immigrant users and Chinese RedNote users. Discussions related to health and wellness are particularly prominent, with keywords such as "cough," "cold," "pain," and "traditional Chinese medicine" reflecting users’ experiences with illness, treatment approaches, and medical concerns. American users tend to share experiences with over-the-counter medications, while terms like "traditional Chinese medicine" appear frequently in Chinese users’ exchanges about traditional remedies.

Food and cooking are also major topics, with terms such as "steamed egg," "chocolate," "tofu pudding," "donut," and "chicken" appearing frequently. Chinese dishes often spark curious questions from overseas users, whereas Western treats like "chocolate" and "donuts" capture the attention of Chinese users.

Interestingly, discussions about household items like "toilet seat" often highlight humorous comparisons of lifestyle habits between regions. Consumption-related topics display distinct regional characteristics: Chinese brands such as "Huawei" and "Xiaomi" are commonly recommended by local users, while international brands are more frequently discussed by overseas users.

Seasonal topics, such as "New Year's Eve dinner" and "Spring Festival," stand out for their cultural resonance—Chinese users typically provide cultural explanations, while overseas users actively engage by sharing their experiences.

\subsubsection{Cultural Topic}
The Cultural Topic encompasses a wide range of user-generated content and reveals how TikTok refugees entering RedNote interact with the platform’s existing Chinese user base through expressions of entertainment interests, foreign cultural imagination, and Chinese regional identity. Popular culture forms a key part of this topic, with users frequently discussing global music trends—such as rock and pop stars like Taylor Swift—as well as fan art, manga, and festive entertainment programs like the Spring Festival Gala. These discussions are often accompanied by humorous expressions (e.g., “haha”) and social gestures such as mutual follows, creating a friendly and relaxed community atmosphere.

In addition, users demonstrate strong interest and imagination toward foreign cultures and Chinese culture, engaging in conversations about countries such as Russia, the United Kingdom, France, South Korea, and India. These discussions involve dimensions of history, language, and media representations—for example, references to the “Soviet Union,” “British style,” “K-dramas,” and the “caste system.” Users often introduce themselves using English or pinyin expressions (e.g., “Hello, I’m from America,” “Montreal”), and sometimes employ historically loaded terms (such as “Xiongnu”) to position foreign users within culturally specific frames.Chinese regional diversity emerges as a common topic of discussion among both Chinese and foreign users. Shared content often includes references to cities and landmarks (e.g., Chengdu, Jiuzhaigou, Guangdong), cultural and natural symbols (e.g., pandas, Sanxingdui, Huanglong), and local cultural elements such as cuisine, dialects (e.g., Cantonese, Guizhou dialect), and internet slang (e.g., “IP address,” “sis,” “emoji”).

As shown above, users from both China and the United States blend global pop culture, foreign cultural perceptions, and localised cultural expressions to construct a flexible and resonant system of reference, which supports identity negotiation and community interaction on the platform.

\begin{figure}[htbp]
    \centering
    \includegraphics[width=\textwidth]{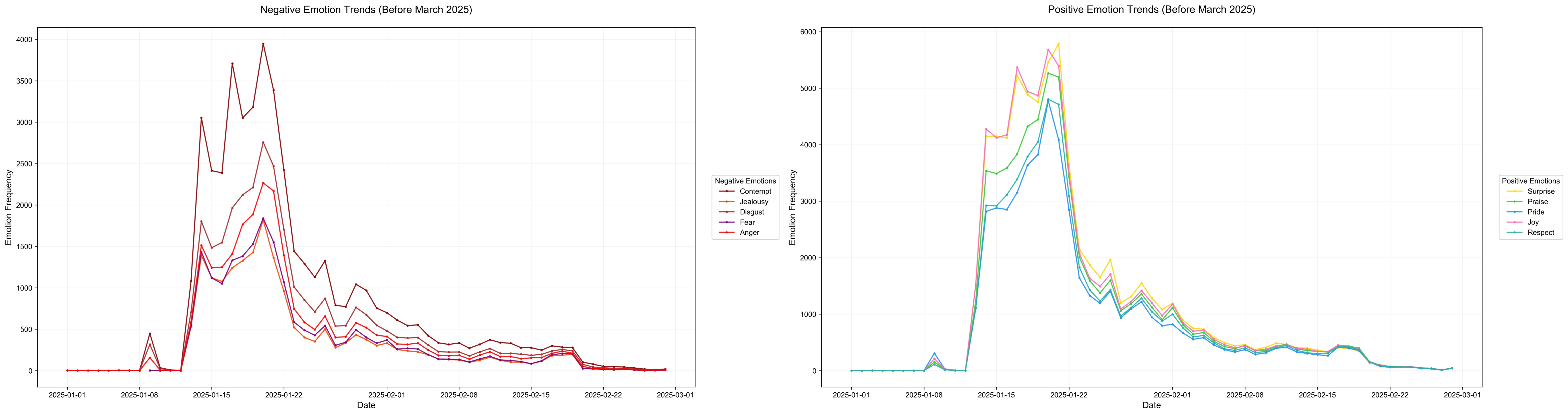}
    \caption{Trends of Positive and Negative Emotions}
    \label{fig:TrendEmotion}
\end{figure}

\subsubsection{Appearance-Related Content}
The Appearance-related content focuses on how Chinese users on RedNote comment on the physical appearance of incoming TikTok refugee users, particularly in response to selfies posted by foreign users. Comments frequently include terms such as “beautiful,” “girl,” and “good-looking,” reflecting a blend of Chinese aesthetic standards and localized expressions. In addition to conventional praise, it is common for Chinese users to post humorous or teasing remarks involving body parts such as the waist, hips, or abs, often conveyed in exaggerated or playful tones. Some comments also reference terms like “Huji” (胡姬), a culturally and historically loaded stereotype used to describe foreign women in a derogatory manner. This suggests that evaluations of American, or more broadly, non-East Asian appearances are not only shaped by globalized beauty ideals but are also deeply influenced by Chinese cultural narratives and linguistic conventions.

\subsubsection{TikTok Refugee Phenomenon}
This topic centers on meta-discursive commentaries about the TikTok refugee phenomenon and its associated labels, revealing users’ critical engagement with content creation logics, platform norms, and the politics of cross-cultural representation. High-frequency terms such as 'truth' , 'staged' , 'plot twist', 'clickbait' and 'algorithm' signal skepticism about authenticity, editing choices, and intentions for attention.
Representative comments like “Is this staged?” “Over-edited,” “Just for traffic,” and “Are you really a refugee?” point to community skepticism regarding the performativity and labeling of TikTok migrants. The term “refugee” itself becomes a discursive flashpoint, with some users accusing it of exaggeration or emotional manipulation, thereby opening up debates around symbolic legitimacy and cultural sensitivity. Some comments also criticize platform dynamics directly, e.g., “The algorithm is weird,” “RedNote has changed,” reflecting discontent toward recommendation logics and anxiety over the platform’s shifting cultural boundaries. These statements move beyond individual videos to engage with the broader governance of content visibility and user norms.

\begin{figure}[htbp]
    \centering
    \includegraphics[width=0.7\textwidth]{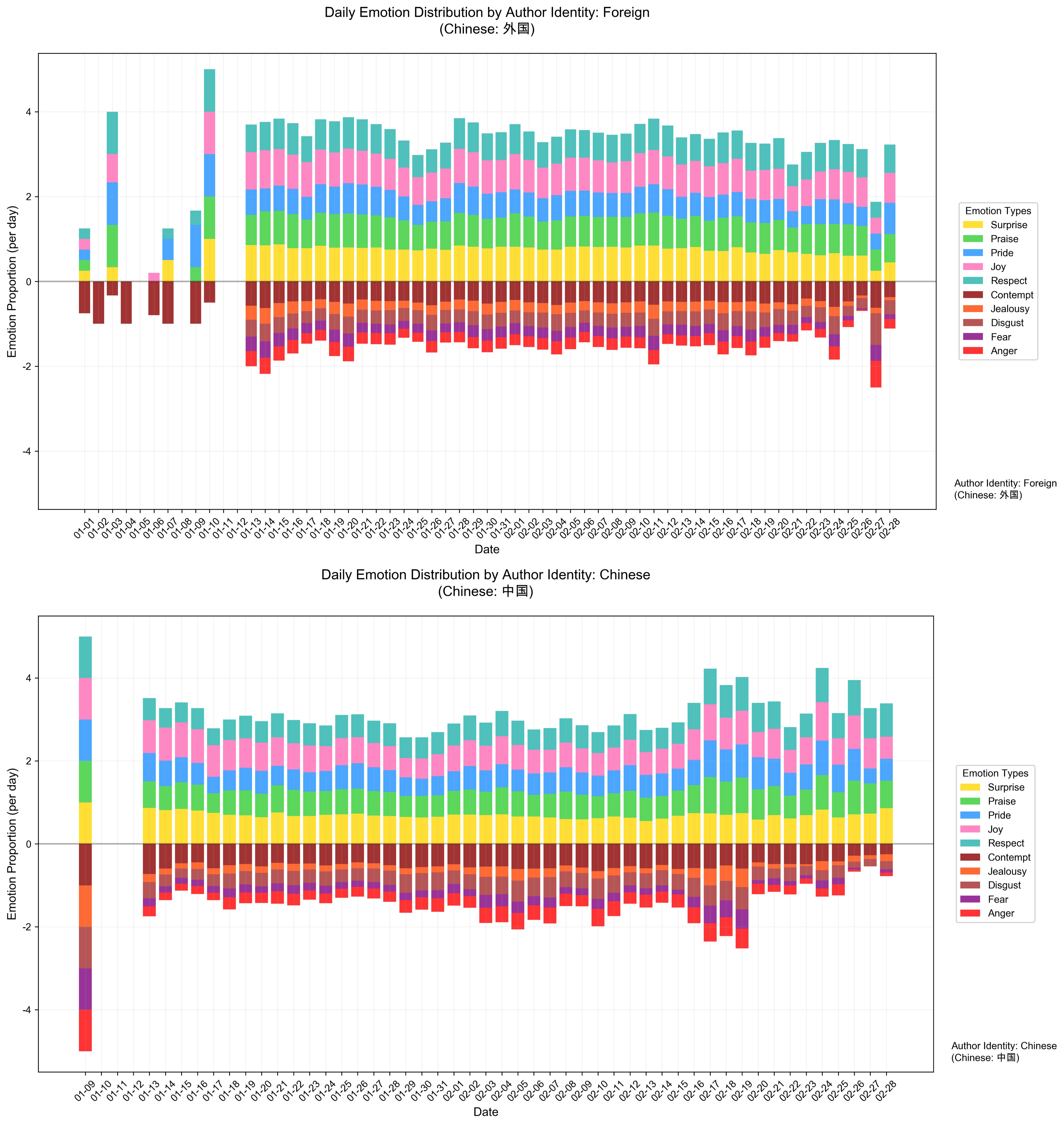}
    \caption{Daily Emotion Distribution of Comments on Notes by Foreign and Chinese Authors}
    \label{fig:DailyEmo_Identity}
\end{figure}

\begin{figure}[htbp]
    \centering
    \begin{subfigure}[b]{0.45\textwidth}
        \centering
        \includegraphics[width=\textwidth]{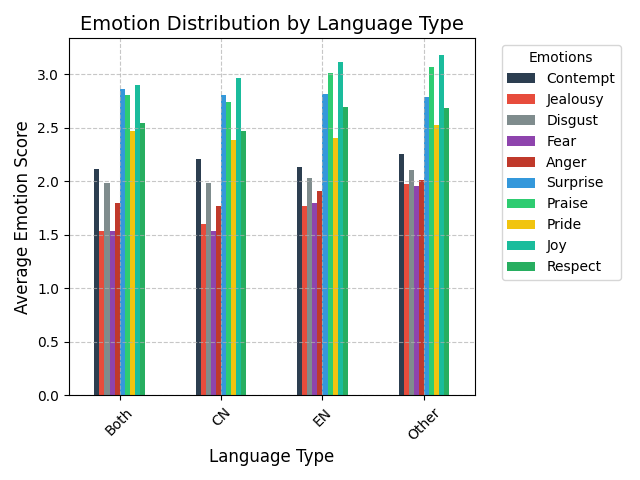}
        \caption{Emotion Distribution of Comments by Language Type}
        \label{fig:emoByLanguage}
    \end{subfigure}
    \hfill
    \begin{subfigure}[b]{0.45\textwidth}
        \centering
        \includegraphics[width=\textwidth]{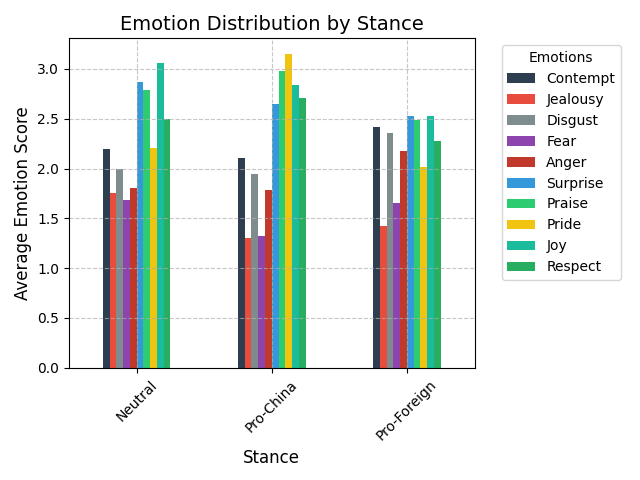}
        \caption{Emotion Distribution of Comments by Stance}
        \label{fig:emoByStance}
    \end{subfigure}
    \caption{Emotion Distribution Analysis}
    \label{fig:main}
\end{figure}

\subsection{Emotion and Stance Distribution}

\subsubsection{Temporal Trends of Emotion}

Figugre \ref{fig:TrendEmotion} illustrate the frequency of negative and positive emotions in comments over time before March 2025. The left graph shows a sharp peak in negative emotions around mid-January, particularly in categories like Anger and Contempt, which then gradually decline. The right graph depicts positive emotions peaking slightly later, with categories like Surprise and Joy showing significant increases before a steady decline. Both graphs indicate a notable fluctuation in emotional expressions during early 2025.

The charts in Figure \ref{fig:DailyEmo_Identity} show daily emotion distribution by identity of note author—foreign and Chinese. Each displays fluctuations in emotions like Surprise, Pride, Joy, and others. Both groups exhibit diverse emotional expressions, with some emotions peaking at different times.

\subsubsection{Across Topic Categories}
Our analysis of emotion and stance distribution across five primary topic categories(The Pet Topic will be ignored here) reveals nuanced patterns in how TikTok Refugee-related discussions unfold on RedNote. The data highlights distinct emotional and attitudinal tendencies depending on the subject matter, with Political topic eliciting stronger negative emotions, while cultural and appearance-related content tends to foster more positive engagement. Below, we break down these trends by topic category.  \\

\textbf{Political Topic}\\
Political discussions surrounding TikTok Refugees exhibit the most polarized emotional responses. Comments under political posts demonstrate heightened levels of contempt (average score within 2.40–2.74) and anger (average score within 2.04–2.83), particularly when the commenter adopts a Pro-China stance (contempt: 2.44, anger: 2.26). Notably, Pro-Foreign commenters under Chinese-authored political posts express the strongest disgust (3.21).  \\

In contrast, neutral commenters exhibit moderated emotions, with contempt averaging 2.00–2.40, indicating that overt political alignment amplifies affective polarization. Positive emotions like pride (average score within 3.08 for Pro-China) and praise (average score within 3.02 for Pro-China) are most pronounced among China-leaning respondents. \\

 \begin{table}[htbp]
        \centering
        \caption{\textbf{Emotion Distribution Across Topic 2  }}
        \adjustbox{width=\textwidth}{
            \begin{tabular}{@{}lcccccc@{}}
                \toprule
                \textbf{Author Identity} &  & \textbf{Foreigner} & \textbf{} & \textbf{} & \textbf{Chinese} & \textbf{} \\ 
                \midrule
                \textbf{Commentor Stance} & \textbf{Pro-Foreign} & \textbf{Pro-China} & \textbf{Neutral} & \textbf{Pro-Foreign} & \textbf{Pro-China} & \textbf{Neutral} \\ 
                \midrule
                \textbf{Num. of Comments} & 98 & 402 & 948 & 121 & 2851 & 1062 \\ 
                \textbf{Account (\%)} & 0.11 & 0.44 & 1.03 & 0.13 & 3.1 & 1.16 \\ 
                \midrule
                \textbf{Average Scores of Negative Emotions} & & & & & & \\ 
                Contempt & 2.663 & 2.438 & 2.401 & 2.736 & 2.427 & 2.005 \\ 
                Jealousy & 1.510 & 1.353 & 1.828 & 1.091 & 1.185 & 1.352 \\ 
                Disgust & 2.684 & 2.400 & 2.187 & 3.215 & 2.321 & 1.907 \\ 
                Fear & 2.000 & 1.632 & 1.978 & 1.603 & 1.335 & 1.374 \\ 
                Anger & 2.500 & 2.259 & 2.043 & 2.835 & 2.202 & 1.761 \\ 
                \midrule
                \textbf{Average Scores of Positive Emotions} & & & & & & \\ 
                Surprise & 2.469 & 2.458 & 2.809 & 2.256 & 2.323 & 2.453 \\ 
                Praise & 2.347 & 2.838 & 2.978 & 2.099 & 3.019 & 2.646 \\ 
                Pride & 2.173 & 3.080 & 2.516 & 1.455 & 3.109 & 2.174 \\ 
                Joy & 2.306 & 2.473 & 2.948 & 1.785 & 2.461 & 2.605 \\ 
                Respect & 2.265 & 2.759 & 2.782 & 2.174 & 2.851 & 2.521 \\ 
        \bottomrule
    \end{tabular}}
\end{table}
%\end{comment}
\textbf{Social Issues}\\
Discussions on societal matters (e.g., economic disparities, social welfare) evoke a mix of moderate negativity and measured optimism. Pro-China commenters display lower contempt (1.63) and anger (1.47) compared to political debates, but Pro-Foreign responses remain critical, with elevated disgust (2.68) and anger (2.43) under Chinese-authored posts.  \\

A striking finding is the high praise (3.38) and pride (3.52) among Pro-China respondents to foreign-authored posts, suggesting that social comparisons often favor China in socioeconomic discussions. Neutral commenters, while less emotionally charged, still lean toward surprise (2.85) and respect (2.79), possibly reflecting curiosity about cross-cultural differences . \\

 \begin{table}[htbp]
        \centering
        \caption{\textbf{Emotion Distribution Across Topic 3  }}
        \adjustbox{width=\textwidth}{
            \begin{tabular}{@{}lcccccc@{}}
                \toprule
                \textbf{Author Identity} &  & \textbf{Foreigner} & \textbf{} & \textbf{} & \textbf{Chinese} & \textbf{} \\ 
                \midrule
                \textbf{Commentor Stance} & \textbf{Pro-Foreign} & \textbf{Pro-China} & \textbf{Neutral} & \textbf{Pro-Foreign} & \textbf{Pro-China} & \textbf{Neutral} \\ 
                \midrule
                \textbf{Num. of Comments} & 304 & 3889 & 4546 & 423 & 1296 & 6005 \\ 
                \textbf{Account (\%)} & 0.33 & 4.23 & 4.94 & 0.46 & 1.41 & 6.53 \\ 
                \midrule
                \textbf{Average Scores of Negative Emotions} & & & & & & \\ 
                Contempt & 2.464 & 1.634 & 2.132 & 2.749 & 2.436 & 2.303 \\ 
                Jealousy & 1.428 & 1.322 & 1.708 & 1.589 & 1.611 & 1.886 \\ 
                Disgust & 2.543 & 1.535 & 1.933 & 2.683 & 2.303 & 2.135 \\ 
                Fear & 1.885 & 1.232 & 1.656 & 1.723 & 1.444 & 1.656 \\ 
                Anger & 2.342 & 1.474 & 1.759 & 2.426 & 1.993 & 1.850 \\ 
                \midrule
                \textbf{Average Scores of Positive Emotions} & & & & & & \\ 
                Surprise & 2.520 & 2.787 & 2.846 & 2.300 & 2.642 & 2.788 \\ 
                Praise & 1.875 & 3.381 & 2.956 & 2.433 & 2.681 & 2.694 \\ 
                Pride & 1.681 & 3.520 & 2.381 & 1.960 & 2.926 & 2.257 \\ 
                Joy & 1.888 & 2.858 & 2.918 & 2.201 & 2.492 & 2.623 \\ 
                Respect & 1.898 & 3.069 & 2.786 & 2.191 & 2.578 & 2.571 \\ 
        \bottomrule
    \end{tabular}}
\end{table}

\textbf{Cultural Topic}\\
Cultural exchanges generate the most balanced emotional landscape. While Pro-Foreign commenters under Chinese posts express strong disgust (2.74), the broader trend is one of cultural appreciation, with Pro-China respondents showing high praise (3.30) and happiness (3.21). Neutral commenters, the largest group (16.29\% of all comments), exhibit moderate positivity (surprise: 2.87, happiness: 3.05), indicating that cultural content fosters less adversarial engagement.\\  

However, appearance-related mockery (e.g., critiques of foreign influencers’ looks) surfaces in Pro-Foreign contempt (2.70) under Chinese posts, hinting at underlying tensions in aesthetic norms.  \\

 \begin{table}[htbp]
        \centering
        \caption{\textbf{Emotion Distribution Across Topic 4  }}
        \adjustbox{width=\textwidth}{
            \begin{tabular}{@{}lcccccc@{}}
                \toprule
                \textbf{Author Identity} &  & \textbf{Foreigner} & \textbf{} & \textbf{} & \textbf{Chinese} & \textbf{} \\ 
                \midrule
                \textbf{Commentor Stance} & \textbf{Pro-Foreign} & \textbf{Pro-China} & \textbf{Neutral} & \textbf{Pro-Foreign} & \textbf{Pro-China} & \textbf{Neutral} \\ 
                \midrule
                \textbf{Num. of Comments} & 383 & 1728 & 8698 & 467 & 3200 & 14975 \\ 
                \textbf{Account (\%)} & 0.42 & 1.88 & 9.46 & 0.51 & 3.48 & 16.29 \\ 
                \midrule
                \textbf{Average Scores of Negative Emotions} & & & & & & \\ 
                Contempt & 2.206 & 1.825 & 2.176 & 2.700 & 2.163 & 2.296 \\ 
                Jealousy & 1.426 & 1.361 & 1.880 & 1.325 & 1.195 & 1.577 \\ 
                Disgust & 2.052 & 1.709 & 2.026 & 2.739 & 1.866 & 1.995 \\ 
                Fear & 1.397 & 1.318 & 1.783 & 1.762 & 1.276 & 1.568 \\ 
                Anger & 1.896 & 1.587 & 1.867 & 2.493 & 1.640 & 1.767 \\ 
                \midrule
                \textbf{Average Scores of Positive Emotions} & & & & & & \\ 
                Surprise & 2.671 & 2.829 & 2.955 & 2.527 & 2.733 & 2.870 \\ 
                Praise & 2.956 & 3.302 & 2.986 & 2.195 & 2.671 & 2.400 \\ 
                Pride & 2.196 & 3.371 & 2.462 & 1.987 & 2.958 & 1.977 \\ 
                Joy & 2.676 & 3.206 & 3.217 & 2.233 & 3.073 & 3.050 \\ 
                Respect & 2.710 & 3.006 & 2.675 & 2.071 & 2.381 & 2.182 \\ 
        \bottomrule
    \end{tabular}}
\end{table}

\textbf{Appearance-Related Content}\\
Posts evaluating physical appearance elicit bifurcated reactions: Pro-China commenters under Chinese-authored content display extreme contempt (3.04) and anger (3.04), likely tied to moral judgments (e.g., perceived vanity). Conversely, foreign-authored appearance posts attract higher praise (3.13 from Pro-Foreign, 3.47 from neutral), suggesting that foreign creators are more often celebrated for their looks.  \\

Neutral commenters dominate this category (8.59\% of all comments) and express strong happiness (3.16) and surprise (2.91), reinforcing the role of appearance-driven content in fostering lighthearted engagement.\\

 \begin{table}[htbp]
        \centering
        \caption{\textbf{Emotion Distribution Across Topic 5  }}
        \adjustbox{width=\textwidth}{
            \begin{tabular}{@{}lcccccc@{}}
                \toprule
                \textbf{Author Identity} &  & \textbf{Foreigner} & \textbf{} & \textbf{} & \textbf{Chinese} & \textbf{} \\ 
                \midrule
                \textbf{Commentor Stance} & \textbf{Pro-Foreign} & \textbf{Pro-China} & \textbf{Neutral} & \textbf{Pro-Foreign} & \textbf{Pro-China} & \textbf{Neutral} \\ 
                \midrule
                \textbf{Num. of Comments} & 98 & 416 & 7897 & 28 & 326 & 1962 \\ 
                \textbf{Account (\%)} & 0.11 & 0.45 & 8.59 & 0.03 & 0.35 & 2.13 \\  
                \midrule
                \textbf{Average Scores of Negative Emotions} & & & & & & \\ 
                Contempt & 2.061 & 1.950 & 2.085 & 2.821 & 3.043 & 1.796 \\ 
                Jealousy & 1.418 & 1.195 & 1.706 & 1.571 & 1.488 & 1.737 \\ 
                Disgust & 1.776 & 1.728 & 1.861 & 2.750 & 2.994 & 1.619 \\ 
                Fear & 1.143 & 1.166 & 1.534 & 1.429 & 1.521 & 1.274 \\ 
                Anger & 1.541 & 1.543 & 1.629 & 2.679 & 3.040 & 1.389 \\
                \midrule
                \textbf{Average Scores of Positive Emotions} & & & & & & \\ 
                Surprise & 2.643 & 2.632 & 2.913 & 2.429 & 2.405 & 2.931 \\ 
                Praise & 3.133 & 2.942 & 2.896 & 2.286 & 1.844 & 3.467 \\ 
                Pride & 1.684 & 2.697 & 1.910 & 1.750 & 2.567 & 2.239 \\ 
                Joy & 2.724 & 2.983 & 3.158 & 2.143 & 1.761 & 3.213 \\ 
                Respect & 2.469 & 2.707 & 2.423 & 2.036 & 1.951 & 2.843 \\ 
        \bottomrule
    \end{tabular}}
\end{table}

\textbf{TikTok Refugee Phenomenon}\\
Meta-discussions about the TikTok Refugee trend itself reveal asymmetric emotional investments. Pro-China commenters exhibit low anger (1.66) but high pride (3.33), framing the phenomenon as a victory for Chinese platforms. In contrast, Pro-Foreign responses are more critical (contempt: 2.15, disgust: 1.97), though neutral commenters—the largest cohort (12.74\%)—lean toward curiosity (surprise: 2.88, happiness: 3.20). \\
 \begin{table}[htbp]
        \centering
        \caption{\textbf{Emotion Distribution Across Topic 6  }}
        \adjustbox{width=\textwidth}{
            \begin{tabular}{@{}lcccccc@{}}
                \toprule
                \textbf{Author Identity} &  & \textbf{Foreigner} & \textbf{} & \textbf{} & \textbf{Chinese} & \textbf{} \\ 
                \midrule
                \textbf{Commentor Stance} & \textbf{Pro-Foreign} & \textbf{Pro-China} & \textbf{Neutral} & \textbf{Pro-Foreign} & \textbf{Pro-China} & \textbf{Neutral} \\ 
                \midrule
                \textbf{Num. of Comments} & 841 & 2248 & 11711 & 219 & 2498 & 10314 \\ 
                \textbf{Account (\%)} & 0.91 & 2.44 & 12.74 & 0.24 & 2.72 & 11.22 \\ 
                \midrule
                \textbf{Average Scores of Negative Emotions} & & & & & & \\ 
                Contempt & 2.151 & 1.891 & 2.283 & 2.411 & 2.135 & 2.090 \\ 
                Jealousy & 1.473 & 1.355 & 1.944 & 1.311 & 1.267 & 1.683 \\ 
                Disgust & 1.969 & 1.770 & 2.114 & 2.384 & 1.983 & 1.957 \\ 
                Fear & 1.605 & 1.346 & 1.927 & 1.744 & 1.447 & 1.746 \\ 
                Anger & 1.856 & 1.658 & 1.964 & 2.224 & 1.747 & 1.789 \\ 
                \midrule
                \textbf{Average Scores of Positive Emotions} & & & & & & \\ 
                Surprise & 2.618 & 2.715 & 2.878 & 2.580 & 2.711 & 2.880 \\ 
                Praise & 2.725 & 3.242 & 2.942 & 2.224 & 2.859 & 2.747 \\ 
                Pride & 2.268 & 3.325 & 2.325 & 1.726 & 2.979 & 2.280 \\ 
                Joy & 3.146 & 3.158 & 3.198 & 2.420 & 2.920 & 3.109 \\ 
                Respect & 2.415 & 2.832 & 2.618 & 2.100 & 2.518 & 2.472 \\ 
        \bottomrule
    \end{tabular}}
\end{table}

\textbf{Positional Distribution Across Topic Categories}\\
The analysis of stance distribution reveals how platform affordances shape ideological alignment across different discussion contexts. Political content (Topic 2) demonstrates the most polarized positioning, with only 53.6\% neutral comments compared to 75-92\% in other categories. Here, pro-China alignment reaches 44.4\% - nearly triple the platform average - while pro-foreign positioning remains minimal (1.9\%), suggesting these threads function as ideological proving grounds where nationalist identities are performed and reinforced.\\

Cultural discussions (Topic 4) present a more nuanced configuration. While maintaining substantial neutrality (75.2\%), these threads show elevated pro-China alignment (23.7\%) that significantly exceeds the TTR phenomenon (15.7\%) and social issue (16.0\%) baselines. This pattern suggests cultural content serves as a primary vehicle for "soft" national identity expression, where celebratory discourse about traditions and achievements facilitates ideological positioning without overt political framing.\\

The appearance category (Topic 5) stands out with both the highest neutrality (92.5\%) and relatively balanced pro-foreign representation (3.8\% vs 1.0-1.9\% elsewhere). This configuration creates what might be termed a "liminal ideological space" - one of few platform contexts where foreign perspectives achieve meaningful visibility without triggering disproportionate nationalist pushback.

Social issue discussions (Topic 3) mirror the TTR phenomenon's (Topic 1) positional distribution (83.1\% vs 83.3\% neutral), but with slightly elevated pro-China alignment (16.0\% vs 15.7\%). This subtle difference may reflect how discussions of healthcare, education and social welfare invite more frequent comparative framing that gently activates national reference points.

%\begin{comment}
 \begin{table}[htbp]
            \centering
            \caption{\textbf{Stance Distribution Across Topics  }}
            \adjustbox{width=\textwidth}{
             \begin{tabular}{@{}llccc@{}}
        \toprule
        Stance        & Topic                               & Num. of Comments & Account (\%) \\ \midrule
        \multirow{5}{*}{\textbf{Pro-Foreign}} & Political Topic      & 219              & 0.24         \\
                       & Social Issues            & 727              & 0.79         \\
                       & Cultural Topic          & 850              & 0.93         \\
                       & Appearance-Related Content & 126              & 0.14         \\
                       & TikTok Refugee Phenomenon & 1060             & 1.15         \\ \midrule
         \multirow{5}{*}{\textbf{Pro-China}}     & Political Topic      & 3253             & 3.54         \\
                       & Social Issues            & 5185             & 5.64         \\
                       & Cultural Topic          & 4928             & 5.36         \\
                       & Appearance-Related Content & 742              & 0.80         \\
                       & TikTok Refugee Phenomenon & 4746             & 5.16         \\ \midrule
        \multirow{5}{*}{\textbf{Neutral}}         & Political Topic      & 2010             & 2.19         \\
                       & Social Issues           & 10551            & 11.47        \\
                       & Cultural Topic          & 23673            & 25.75        \\
                       & Appearance-Related Content & 9859             & 10.72        \\
                       & TikTok Refugee Phenomenon & 22025            & 23.96        \\ \bottomrule
    \end{tabular}}
\end{table}

\subsubsection{Across Stance Categories}

Our stance-specific analysis reveals systematic variations in emotional expression across \textit{Neutral}, \textit{Pro-China}, and \textit{Pro-Foreign} commenters, with distinct affective profiles emerging for each group. These findings underscore how ideological alignment shapes not only the content but also the \textit{emotional tenor} of responses to TikTok Refugee-related discussions on RedNote. Refer to Figure \ref{fig:emoByStance} for details on the emotion distribution by comment stance.\\

\textit{Neutral Commenters: Measured Engagement with Leaning Positivity}\\

Neutral commenters, constituting the largest stance group, exhibit a \textit{moderate emotional profile}, balancing mild negativity with stronger positive affect. Their \textit{contempt (2.20)} and \textit{anger (1.80)} scores are the lowest among the three stances, suggesting deliberate restraint in adversarial positioning. However, they display \textit{pronounced positive emotions}, particularly \textit{joy (3.06)} and \textit{surprise (2.87)}, indicating curiosity-driven engagement rather than ideological confrontation. Notably, their \textit{pride (2.21)} and \textit{respect (2.50)} levels are intermediate—higher than Pro-Foreign but lower than Pro-China commenters.\\

\textbf{Pro-China Commenters: Assertive Positivity with Selective Negativity}\\

Pro-China respondents demonstrate a \textit{dual emotional strategy}: they combine \textit{low-intensity negativity} (contempt: 2.11, anger: 1.79) with \textit{exceptionally high pride (3.15)} and \textit{praise (2.98)}, reinforcing a narrative of \textit{national superiority}. Their \textit{fear (1.32)} and \textit{disgust (1.95)} scores are the lowest of all groups, suggesting that their stance is less reactive to perceived threats and more focused on \textit{affirmative self-presentation}. \\

\textbf{Pro-Foreign Commenters: Critical Negativity with Muted Positivity}\\

In stark contrast, Pro-Foreign commenters exhibit the \textit{strongest negative emotions}, including \textit{highest contempt (2.42)}, \textit{disgust (2.36)}, and \textit{anger (2.17)}. This pattern suggests a \textit{defensive posture}, likely in response to perceived \textit{anti-foreign sentiment} or \textit{platform bia}s. Their \textit{positive emotions are subdued}: \textit{joy (2.52)} and \textit{respect (2.28)} are significantly lower than Neutral or Pro-China groups, and their \textit{pride (2.02)} is the weakest, reflecting either \textit{alienation} or \textit{strategic avoidance} of nationalist rhetoric. Surprisingly, their \textit{surprise (2.53)} is also lower than Neutral users, implying less exploratory engagement and more \textit{targeted criticism}.

\begin{figure}[htbp]
    \centering
    \includegraphics[width=0.8\textwidth]{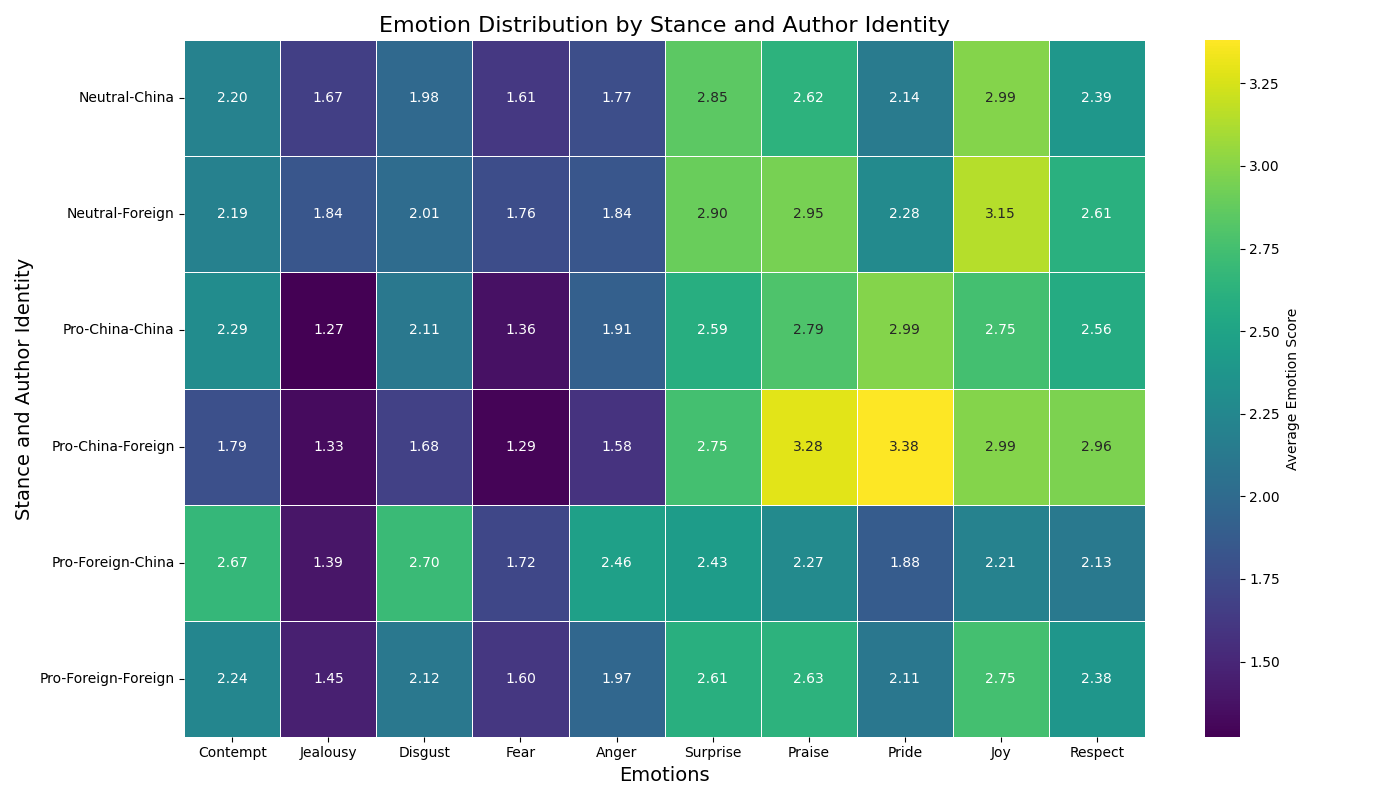}
    \caption{Emotion Distribution Heatmap by Comment Stance and Note Author Identity}
    \label{fig:heatmap}
\end{figure}

Figure \ref{fig:heatmap} further visualizes the correlation between emotion, stance, and the identity of note authors. Emotions like Surprise, Praise, and Joy generally have higher scores, indicating more positive sentiment, while emotions like Contempt, Jealousy, and Fear have lower scores.

\begin{table}[h]
    \centering
    \caption{Emotion Distribution Across Stance}
    \adjustbox{width=\textwidth}{
    \begin{tabular}{@{}lcccccccccc@{}}
        \toprule
         & \multicolumn{5}{c}{\textbf{Average Scores of Negative Emotions}} & \multicolumn{5}{c}{\textbf{Average Scores of Positive Emotions}}\\
        \textbf{Stance} & \textbf{Contempt} & \textbf{Jealousy} & \textbf{Disgust} & \textbf{Fear} & \textbf{Anger} & \textbf{Surprise} & \textbf{Praise} & \textbf{Pride} & \textbf{Joy} & \textbf{Respect} \\ 
        \midrule
        Neutral & 2.197 & 1.750 & 1.997 & 1.685 & 1.800 & 2.872 & 2.784 & 2.206 & 3.063 & 2.498 \\ 
        Pro-China & 2.106 & 1.301 & 1.949 & 1.324 & 1.786 & 2.644 & 2.976 & 3.150 & 2.840 & 2.712 \\ 
        Pro-Foreign & 2.422 & 1.428 & 2.361 & 1.652 & 2.172 & 2.532 & 2.484 & 2.019 & 2.524 & 2.279 \\ 
        \bottomrule
    \end{tabular}}
\end{table}

\begin{figure}[htbp]
    \centering
    \includegraphics[width=0.95\textwidth]{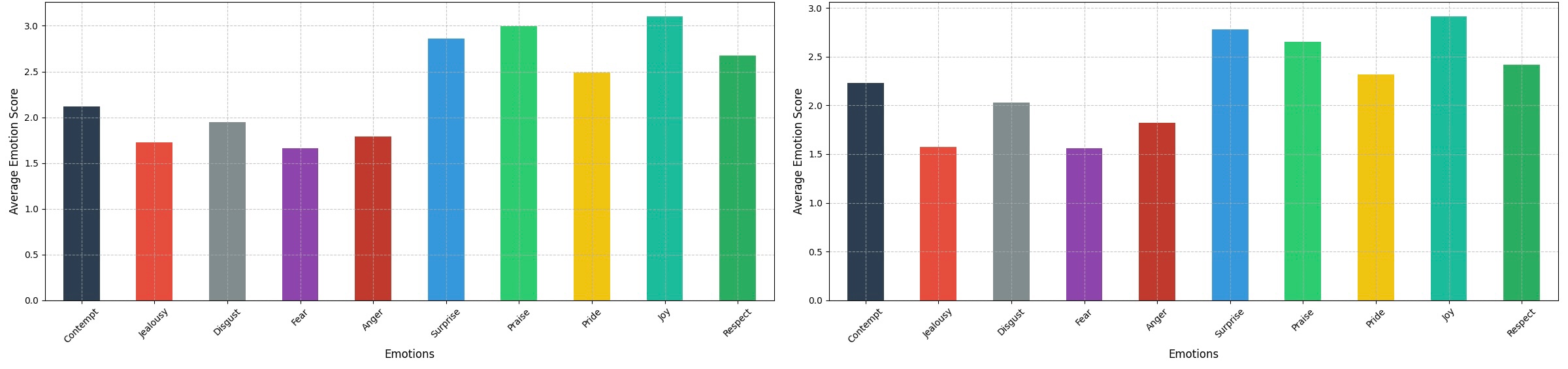}
    \caption{Emotion Distribution of All Comments on Notes by Foreign (Left) and Chinese (Right) Authors}
    \label{fig:EmotionByIdentity}
\end{figure}

\section{Discussions}
\subsection{TikTok Refugees' Cultural Engagement and Emotional Resonance on RedNote}
In mid-January 2025, as the U.S. Supreme Court began its judicial review of the \enquote{Protecting Americans from Foreign Adversary Controlled Applications Act}, TikTok faced a potential nationwide ban. This prompted a wave of American creators to migrate to RedNote, identifying themselves as \enquote{TikTok Refugees} (\cite{Liu2025}). This migration reflects Lee’s push–pull model: TikTok’s policy uncertainty and algorithm transparency issues served as major push factors (\cite{Eslami2015}), while RedNote attracted users with its middle-class lifestyle branding, hybrid image–video interface, and traffic-boosting policies for new users.

These factors may explain why RedNote quickly became a lucky alternative to TikTok. The app rose to popularity beyond China, especially in North America, where users were pushed by TikTok’s uncertainties and pulled by RedNote’s user-friendly design and support for newcomers.

At first, TikTok refugees posted in a typical ``cute pets \& bilingual greeting'' style—showing cats (mostly) with captions like \enquote{Hello from America} or \enquote{Nihao} (\cite{EastIsRead2025}). This aligns with social identity theory’s minimal group paradigm, where shared interests—even minimal ones like pet appreciation—can trigger a sense of temporary belonging (\cite{TajfelTurner1979}). Hashtags like \#TikTokRefugee and \#CatsTax boosted content visibility, and RedNote’s interest-based recommendation system further amplified reach (\cite{Pariser2011}). The ``cat tax'' trend quickly became a cross-cultural interaction hub and one of the platform’s most viral topics, flooding the app with foreign pet posts—cats, dogs, even horses.

Although RedNote filters sensitive political content (\cite{Wu2024}), the political origins of this migration made geopolitics unavoidable. Comments frequently debated the TikTok ban, international relations, and free speech. Users expressed stances through annotated maps, references to official statements, and hashtags like \#SanctionUSImports, showing how politics permeated even lifestyle platforms (\cite{DFRLab2025}).

As the posts multiplied, comment sections expanded from praising pets (\enquote{So healing}, \enquote{Too cute}) to discussions about appearance and social structures. Chinese users and TikTok refugees naturally shifted from pet talk to comparing daily life and media narratives in a kind of ``US–China life audit.'' Even with RedNote’s content filtering (e.g., blocking \#SanctionUSImports), the political roots of this migration allowed geopolitical issues to seep into user interactions. Through symbolic cues like historical references or healthcare cost comparisons (China’s 98\% medical coverage vs.\ America’s 89\%), users engaged in subtle political dialogue, becoming grassroots evaluations of national governance. These ``lifestyle translations'' helped avoid overt conflict while challenging long-held cultural stereotypes.

But this doesn’t mean a new era of openness. In appearance-related posts, Chinese users often praised foreigners as \enquote{beautiful} or \enquote{good-looking}, but sometimes included jokes rooted in outdated stereotypes (e.g., \enquote{Huji}). Emotion analysis showed higher levels of contempt and anger in comments about \enquote{foreign} topics, and elevated pride when discussing \enquote{China}. This emotional asymmetry aligns with in-group favoritism and out-group derogation in social identity theory (\cite{TajfelTurner1979}).

Such emotional polarization is amplified by platform algorithms. Even as RedNote blocks sensitive hashtags, its recommendation engine favors controversial topics—pet care cost comparisons got three times the usual distribution—creating algorithmic pathways for emotional mobilization. To gain visibility, users simplify complex issues into ``China vs.\ U.S.'' data contrasts. While efficient, this narrative simplification fuels cognitive bias. When algorithms repeatedly favor ``China wins'' content, nationalism is reinforced, creating an illusion of ``data justice.''

Compared to the 2022 Twitter-to-Mastodon migration, the TikTok refugee wave shows key differences. Mastodon users were driven by tech idealism and anti-centralization values, making their move a political statement. TikTok users, by contrast, were motivated by practical risk avoidance, reflecting a functional replacement logic (\cite{Lee1966}).

This also shaped their content strategies: Mastodon users focused on platform governance, aiming to build alternative public spheres. RedNote migrants embedded political topics into everyday content—e.g., pet costs as metaphors for health system differences—enabling subtle political expression under algorithmic mediation (\cite{Gillespie2018}). Social capital transfer also differed. Twitter users tried to replicate strong ties with tools like Debirdify, but lost connections due to decentralization. TikTok refugees abandoned fanbase migration, instead rebuilding weak-tie networks via hashtags like \#HelloFromAmerica, forming temporary belonging mediated by algorithms (\cite{TajfelTurner1979}).

Ultimately, these differences reveal core tensions in digital migration. As platform control shifts from visible censorship to invisible algorithmic governance (\cite{vanDijck2013}), user resistance evolves—from tech empowerment to algorithmic co-adaptation. The Mastodon case exposed the governance limits of decentralization, while the RedNote case revealed the ethical paradox of algorithm-mediated cross-cultural interaction: lifestyle content may look harmless, but by rewarding stereotypes, the system reproduces symbolic violence while masking deeper conflict (\cite{Bourdieu1984}).

\subsection{Pro-China Behaviors and Affective Structures of TikTok "Refugees"}
Unlike explicit moderation strategies common on Western platforms, RedNote relies heavily on emotional feedback from the community to implicitly regulate content. Users collectively govern through affective responses—such as sarcasm, contempt, and approval—rather than relying on explicit content removals or overt moderation. This form of emotional governance aligns closely with what Schoenebeck and Blackwell (2020) conceptualize as "normative regulation," where implicit rules are learned and reinforced through ongoing interactions rather than formal moderation systems. However, such implicit governance through community feedback can inadvertently strengthen cultural divides, as emotional policing often reflects and reinforces existing cultural prejudices and nationalist sentiments rather than fostering genuine cross-cultural dialogue.

Through a quantitative analysis of discussions by “TikTok refugees” on RedNote, we find that in comments on China-positive posts—especially those addressing Chinese social governance and cultural narratives—pro-China commenters exhibit intense positive emotional engagement. For example, under political themes, their mean “pride” score reaches 3.08 (SD = 0.41), which is 32.6\% higher than that of neutral commenters; in cultural themes, their mean “praise” (3.30) and “joy” (3.21) scores significantly exceed the platform benchmark of 2.85.(\cite{BCPublication2025})

The pro-China tendency of TikTok refugees on RedNote is not accidental but arises from the interplay of selective self-presentation, algorithmic amplification, and platform governance. First, a large wave of “refugee” users poured into RedNote in a short span, self-tagging with \#TikTokRefugee and \#HelloFromAmerica while actively sharing Spring Festival customs, Hanfu experiences, and cost-of-living comparisons between China and the U.S., thus framing Chinese culture and social systems as worthy of admiration and identification.(\cite{Reuters2025a, APNews2025a, Reuters2025b}) This “reverse political statement” served both as implicit protest against the U.S. TikTok ban and as a strategy to avoid censorship of sensitive topics.(\cite{DFRLab2025, Eslami2015}) Under the platform’s algorithmic incentives, “harmless” content such as pet photos and daily life comparisons gained greater recommendation and exposure, thereby continually reinforcing the visibility of pro-China discourse.(\cite{Pariser2011, BCPublication2025})

Emotion distribution data show that pro-China commenters in political discussions simultaneously display strong positive and negative emotions: “pride” (M ≈ 3.08) and “praise” (M ≈ 3.02) are high, while “contempt” (M ≈ 2.44) and “anger” (M ≈ 2.26) are also significantly higher than in neutral and pro-foreign groups.(\cite{BCPublication2025}) This bipolar emotional pattern confirms the social identity theory mechanism of in-group favoring and out-group derogation: pro-China users consolidate group boundaries through contempt toward dissenters, while reinforcing Chinese identity with pride and praise.(\cite{Tajfel1979}) By contrast, neutral users participate less in political topics, maintaining moderate emotional scores and often adopting a “strategic silence” to avoid confrontation.

In social‐issue discussions, pro-China commenters give higher “praise” (M ≈ 3.28) and “pride” (M ≈ 3.38) scores to foreign authors’ posts, indicating that socioeconomic comparisons are frequently used to highlight the superiority of Chinese systems. When U.S. users disclose high tuition and medical debt, Chinese users respond with notes comparing China’s lower living costs. This “China–U.S. reconciliation” not only satisfies cross-cultural curiosity but, aided by algorithmic boosts, gradually forms a collective narrative that further cements pro-China perceptions.(\cite{APNews2025a})

Compared to political and social issues, cultural and appearance topics serve to mitigate conflict and foster resonance. In cultural threads, pro-China users exhibit high “praise” (3.30) and “joy” (3.21), while neutral users show “surprise” (2.87) and “joy” (3.05), reflecting genuine curiosity about Chinese traditions and lifestyles.(\cite{ExpressNews2025}) However, in appearance discussions, foreign (especially white) creators receive higher praise from neutral and pro-China commenters (pro-foreign M = 3.13; neutral M = 3.47), whereas Chinese creators’ photos attract stronger negative emotions (“contempt” M = 3.04; “anger” M = 3.04).(\cite{OurStudy2025}) Some users even use the pejorative historical term “胡姬” to refer to non-native Chinese speakers—a form of “objectification” that treats them as consumable objects of critique.

Platform governance plays a dual role: by blocking sensitive tags (e.g.\ \#SanctionUSImports) and adjusting algorithmic weights, RedNote creates a “buffer zone” for political topics, preventing large-scale conflict.(\cite{Wu2024}) Simultaneously, the continuous promotion of “harmless” life-style content drives a “pet socializing–traffic incentive–content reproduction” loop, which objectively shapes a space conducive to pro-China expression. This “algorithmic collusion” mechanism not only elevates the visibility of “positive energy” content but also suppresses critical discussion of Chinese culture and systems.

In sum, the pro-China behaviors of TikTok refugees on RedNote arise from the combined effects of push–pull migration motives, social identity dynamics, symbolic violence reproduction, and algorithmic governance. This multi-dimensional perspective deepens our understanding of how emotions and stances are generated in digital migration contexts, and it cautions platform administrators to guard against inadvertently amplifying cultural biases and social polarization in the pursuit of traffic and “positive energy.”

\subsection{Identity and Emotion in Social Media Spaces}
The findings of this study demonstrate that, compared to foreign commenters, Chinese users on RedNote exhibit significantly higher emotional intensity across negative affective dimensions, including contempt, disgust, fear, and anger. Chinese users also express substantially more surprise and pride, whereas foreign users show higher levels of respect and admiration.

Such differences not only reflect the structural distribution of user groups on the platform but also reveal the construction of cultural boundaries in digital spaces. As research in intercultural communication has emphasized, language and cultural symbols are not merely tools of information exchange but function as markers of identity and belonging (\cite{Heang2024}).

Linguistic difference constitutes a primary axis of identity. On RedNote, Chinese-speaking users function as “platform natives,” occupying a dominant position in terms of linguistic usage, interactional norms, and cultural values. In contrast, English-speaking users may be considered “digital migrants,” who often struggle to fully integrate into the mainstream communicative environment. This asymmetry in language and cultural fluency reinforces the boundaries between in-groups and out-groups, making Chinese users more prone to defensive emotions and cultural anxiety when encountering comments that carry “otherness.” Linguistic identity also maps onto the majoritarian/minoritarian distinction on the platform, with English users forming a numerical minority. Prior research on minority expression strategies suggests that minority groups often exhibit higher activity levels on social media and adopt more affiliative emotional styles (\cite{Correa2010}), a pattern consistent with our findings. However, most existing studies have focused on institutionally or perceptually marginalized groups. For instance, one study based on the social identity framework found that perceived discrimination among Latinx users significantly increased their willingness to express themselves online (\cite{Velasquez2020}). Another study on sexual minority adolescents showed that they tend to accumulate less bonding capital, have narrower networks, and experience greater social isolation on social media (\cite{Charmaraman2021}). Distinct from such studies, our research draws attention to a form of relative minority status produced within the platform space—not grounded in institutional exclusion or perceived discrimination, but emerging from user configurations and behavioral patterns. English-speaking users on RedNote are not sociopolitical minorities in the conventional sense; rather, they constitute a statistical minority who nonetheless show stronger tendencies toward the expression of positive emotions.

The positioning of English users as "refugees" on the platform resonates with prior scholarship. Putnam’s (2000) distinction between “bonding” and “bridging” forms of social capital offers a useful framework to understand differential emotional expressions. Chinese users are more likely to form bonding capital through homophilous interactions, reinforcing shared national identities and affective communities (\cite{Putnam2000}). English-speaking users, by contrast, face difficulties in establishing bridging capital due to language barriers and marginal group positioning. Lacking interactional affiliation, their expressions tend to be more neutral or external in perspective. This structural asymmetry not only produces diverging tonalities in comments but also contributes to a tendency among Chinese users to perceive foreign remarks as implicit evaluations of domestic values, thereby triggering defensive emotional responses.

National identity, represented by linguistic affiliation, constitutes a second critical axis of identity. Previous research has identified nationalism as a key factor in shaping emotional discourse on Chinese social media. Zhang (2024) argues that digital nationalism is increasingly pervasive in China (\cite{Zhang2024}), while Schneider (2018) (\cite{Schneider2018}) suggests that it functions not only as a conduit for state ideology but also as a vehicle for young people to express national pride and political agency. The intensification of such sentiments may encourage the rejection of foreign cultures and perspectives, which manifest on social media as heightened expressions of negative affect. The existence of Han-centric nationalism, for instance, has contributed to online hostility toward Muslims on Weibo (\cite{Liu2024}). Moreover, the mechanism of social comparison is also at work. According to Tajfel and Turner’s (1979) social identity theory, individuals define themselves through intergroup comparisons, generating in-group favoritism and out-group derogation. (\cite{TajfelTurner1979}) On a predominantly Chinese-language platform like RedNote, English-speaking users are often construed as out-group members and thus become targets of exclusionary or negative affect. Such identity-based and symbolic boundary work ultimately produces a sense of national superiority among Chinese users.

Within the category of "contempt," this study finds that such emotion is not limited to explicitly political content but also emerges in lifestyle-related discourse. Figure \ref{fig:emoByLanguage} presents the emotion distribution accross different language users. For example, the comment “Many Chinese girls were attracted by this tattered wig and ended up going with black people” reflects Han-centric ethnocentrism, embodying exclusionary attitudes and resentment toward perceived foreign admiration. Xenophobia also surfaces in expressions of avoidance, such as “It seems they have already made this place their home; we can go back to preparing for Spring Festival,” which signals diminished willingness to engage with English-language users. From the perspective of the sociology of emotion, this pattern helps to explain why Chinese users are more inclined to express “pride” and “surprise,” whereas foreign users more frequently adopt affective modes such as “respect” and “admiration.” Pride and surprise are often associated with collective national achievements or sudden displays of cultural and technological prowess (e.g., high-speed rail, AI, traditional attire), and are characteristic of “emotional communities” (\cite{Scheer2012}) closely aligned with state narratives. In contrast, respect and admiration are more typical of external observers acknowledging foreign accomplishments—more neutral, evaluative, and reasoned in emotional tone. This form of “emotional displacement” illustrates the tight coupling between identity position and emotional expression.

In summary, this study reveals a structural coupling mechanism between identity positions and emotional expression on social media platforms. Emotion, we argue, is not merely a spontaneous psychological state, but rather a socially structured practice that is shaped by platform architecture, identity perception, and cultural power dynamics. This mechanism can be conceptualized as follows: Platform structures encode identity perception, which in turn shapes group positioning and affects both the directionality and content of emotional expression. This framework helps illuminate how linguistic and cultural identities are constructed and reproduced through affective communication in globalized digital spaces, offering theoretical insight for future research on platform governance, cross-cultural interaction, and affective polarization.

\subsection{Limitation}
Our study offers empirical insights into the emotion–identity dynamics of TikTok Refugees on RedNote, but it is not without limitations. First, our dataset is limited to a three-month window (February to April 2025), capturing only the early phase of cross-cultural encounters. While this period coincides with heightened platform activity following TikTok’s ban, user attitudes and emotional patterns may evolve over time. Longitudinal analysis would be valuable to trace the development of identity negotiation and affective feedback loops beyond the initial wave of interactions.

Second, although we implemented a rigorous classification pipeline using large language models (LLMs) for emotion and stance detection, these outputs remain sensitive to prompt framing and contextual ambiguity. While we designed a multi-layer contextual framework to enhance interpretability, LLMs may still misread sarcasm, irony, or culturally specific expressions—especially in multi-language or code-switched comments. Future work could integrate human-in-the-loop validation or hybrid coding strategies to improve reliability.

Third, our operationalization of user identity relies on indirect indicators such as IP location, usernames, and profile tags. While effective in distinguishing broad cultural backgrounds, this method may overlook diasporic users, multilingual actors, or users who deliberately obfuscate their identity. This limitation could be addressed by incorporating richer digital ethnographic methods or self-reported user demographics, where available.

Finally, our focus on RedNote as a single case study of a region-specific lifestyle platform limits the generalizability of our findings. While RedNote offers a unique lens into Chinese platform cultures and algorithmic governance, comparative analysis with other hybrid platforms (e.g., Bilibili, WeChat Channels, or Mastodon) could further contextualize our claims and test the robustness of the affective governance framework we propose.

\section{Conclusion}
This study explored the emotion–identity dynamics of TtR navigating a Chinese platform environment shaped by culturally specific interaction norms and informal community expectations. Through large-scale sentiment analysis and topic modeling of 403,054 comments on RedNote, we examined how foreign users’ posts received different emotional and symbolic responses depending on topic and tone. We found that cross-cultural interactions were shaped less by official moderation and more by users’ collective reactions—such as praise, correction, sarcasm, and shifts in topic—that signaled what kinds of emotional expression were considered acceptable.

By analyzing RedNote as a case of culturally embedded digital discourse, we show how national identity becomes visible and reinforced through repeated emotional patterns. Chinese users frequently expressed pride and anger in response to political content, while foreign users relied on humor, cultural curiosity, or lifestyle topics to engage more safely. These patterns reflect a shared understanding of platform-appropriate emotion and tone, which structured how visibility and belonging were managed in public interactions.

Our work contributes to CSCW by highlighting how emotional expression is shaped by social expectations and interactional context, especially in identity-relevant and cross-cultural environments. We build on prior work on social norms by showing how emotional tone can serve as a gatekeeping mechanism—determining not only whether a message is accepted, but also how it must be framed to be heard. Emotional interactions, we argue, are central to how users negotiate legitimacy, safety, and inclusion in transnational digital publics.

These insights offer implications for platform design, especially in multilingual or multi-regional contexts. Supporting more equitable cross-cultural engagement requires not only content translation, but also sensitivity to emotional tone, cultural references, and perceived norms of participation. As global platforms continue to host politically and culturally diverse interactions, understanding how emotional dynamics shape user experience will remain critical to building more inclusive digital spaces.
%Bibliography
\bibliographystyle{ACM-Reference-Format}
\spacingset{1}
\bibliography{manuscript}

\end{CJK*}
\end{document}